\documentclass[twocolumn]{aastex63}

\newcommand{\filter}[1]{\mbox{\it #1\/}}              

\newcommand\HH{H72.97-69.39}
\received{\today}
\revised{}
\accepted{}
\submitjournal{ApJ}

\shorttitle{N79 in the LMC}
\shortauthors{Andersen et al.}
\graphicspath{{./}{figures/}}

\begin{document}

\title{The stellar content of H72.97-69.39, a potential super star cluster in the making}

\correspondingauthor{Morten Andersen}
\email{manderse@gemini.edu}

\author[0000-0002-5306-40890]{Morten Andersen }
\affiliation{Gemini Observatory, NSF’s National Optical-Infrared Astronomy Research Laboratory, Casilla 603, La Serena, Chile}

\author{Hans Zinnecker}
\affiliation{Universidad Autonoma de Chile, Avda Pedro de Valdivia 425, Providencia, Santiago de Chile.}

\author[0000-0002-2954-8622]{Alec S. Hirschauer}
\affiliation{Space Telescope Science Institute, 3700 San Martin Drive, Baltimore, MD 21218, USA}

\author{Omnarayani Nayak}
\affiliation{Space Telescope Science Institute, 3700 San Martin Drive, Baltimore, MD 21218, USA}


\author{Margaret Meixner}
\affiliation{Department of Physics \& Astronomy, Johns Hopkins University, 3400 N. Charles Street, Baltimore, MD 21218, USA}
\affiliation{Space Telescope Science Institute, 3700 San Martin Drive, Baltimore, MD 21218, USA}



\begin{abstract}
Young Massive Clusters (YMCs) and Super Star Clusters (SSCs) represent an extreme mode of star formation. 
Far-infrared imaging of the Magellanic Clouds has identified one potential embedded SSC,  HSO BMHERICC J72.971176-69.391112 (\HH\ in short), in the south-west outskirts of the Large Magellanic Cloud. 
We present Gemini Flamingos 2 and GSAOI near-infrared imaging of a 3\arcmin$\times$3\arcmin\ region around  \HH\  in order to characterize the stellar content of the cluster. 
The stellar content is probed down to 1.5 M$_\odot$. 
We find substantial dust extinction across the cluster region, extending up to  $A_{K}$ of 3. 
Deeply embedded stars are associated with  ALMA-detected molecular gas   suggesting that star formation is ongoing. 
The high spatial resolution of the GSAOI data allows identification of  the central massive object associated with the $^{13}CO$ ALMA observations and to detect fainter low-mass stars around the H30$\alpha$ ALMA source. 
The morphology of the molecular gas and the nebulosity from adjacent star formation suggest they have interacted covering a region of several pc. 
{ The total stellar content in the cluster is estimated from the intermediate-- and high--mass stellar content to be at least 10000 M$_\odot$, less than R136 with up to 100 000 M$_\odot$ within 4.7 pc radius, but places it in the regime of a super star cluster. 
Based on the extinction determination of individual stars we estimate a molecular gas mass in the vicinity of \HH\ of 6600 M$_\odot$, suggesting more star formation can be  expected. }

\end{abstract}



\section{Introduction} 
{ The formation and early evolution of massive star clusters, those containing large numbers of early type stars, is currently poorly understood but is fundamental for our understanding of star formation in general and massive star formation in particular \citep[e.g.][]{zinnecker,adamo}. 
The lack of knowledge  is largely due to young massive clusters (YMCs) being statistically rare and thus typically far away,  making it difficult to spatially resolve these clusters into the individual stellar components. }
{ Although there are several very massive clusters in the Milky Way and nearby galaxies, they are typically mature in the sense that they have already expelled most of their natal molecular gas (e.g. Westerlund 1 \citep{andersen09}, Arches \citep{hosek}, and NGC 3603 \citep{stolte_ngc3603} in the Milky Way and R136 \citep{andersen09} within the 30 Doradus star forming complex in the Large Magellanic Cloud (LMC) and NGC604 in M33 \citep{hunter_604}. |}
{ Thus, star formation has mostly ceased in these systems. 
Furthermore, the clusters are much older than the crossing time and will already be dynamically evolved with any sub-structure at formation erased.} 
{
As part of the HERITAGE Herschel program \citep{meixner} it was found that the N79 { star-forming complex} in the LMC, spanning over 500 pc, has twice the star formation rate of 30 Doradus \citep{ochsendorf}. 
Within the region is the very luminous far-infrared {\it Herschel} point source HSO BMHERICC J72.971176-69.391112}, for short \HH, with an integrated luminosity of $2.2\cdot10^6$ L$_\odot$ \citep{seale,ochsendorf}{, located within the HII  region IC2111}. 
{ Although the bolometric luminosity  is less than that of  R136 \citep{malumuth}, it is one of the most luminous star forming clusters known in the LMC and due to the high far-infrared luminosity is suggestive of being a super star cluster in the making \citep{ochsendorf}}. 
Its existence within the LMC makes the target particularly interesting since it allows for studying massive star cluster formation and evolution in a sub-solar metallicity environment.  
Furthermore star clusters in the LMC are less affected by field star contamination than clusters in the Galactic plane which allows for a more straightforward analysis of the stellar population in the region. 

 Little is currently known about the stellar content of the region around \HH . \citet{indebetouw} had earlier used Australia Telescope Compact Array (ATCA)  measured flux densities at 3 and 6 cm to estimate an ionizing flux from the region now known as  \HH\  consistent with an O5 star, assuming optically thin bremsstrahlung.  
However, the total stellar content in the region is not known due to the limited depth of the previous near-infrared  observations shown in \citet{nayak19}. 
ALMA line observations showed a complex of filaments feeding into the central region \citep{nayak19} which can provide additional mass to the region in the future. 
In terms of directly detecting the stellar content, no deep high spatial resolution imaging of the region has so far been performed. 
The extinction is expected to be high in the region and near-infrared observations are necessary.

Here we present near-infrared  Gemini Flamingos 2 and Multi-Conjugate Adaptive Optics  (MCAO) GeMS/GSAOI observations of \HH\  to probe the intermediate-- and high-- mass stellar content of the region. 
The larger field of view of Flamingos 2 provides the larger glimpse of the region and allows calibration of the deeper GeMS/GSAOI data that, through the higher spatial resolution possible with MCAO, can resolve the individual stars in \HH . 
Previous observations of massive star forming regions in the LMC have shown that adaptive optics and in particular multiconjugate adaptive optics allows for resolution of  the individual stars \citep{brandl96,bernard}. 

The paper is structured as follows. 
In Section 2 the observations are presented together with the data reduction and source extraction and photometry. 
Section 3 presents the basic observational findings. Color-magnitude diagrams (CMDs) are presented and different sub-regions within the observed field are described. 
In Section 4 the stellar content is analyzed using pre-main sequence evolutionary models. 
We derive the total mass of the young stellar population and discuss the near-infrared observations in the context of the previous ALMA and Herschel observations. 
Finally we summarize our findings in Section 5. 
 
\section{Data and data reduction} 
N79 has been imaged using the natural seeing infrared imager Flamingos 2 and the MCAO assisted imager GSAOI, both mounted on Gemini South. 
Below we describe the data reduction and calibration of the data. 

\subsection{Flamingos 2 imaging}
As an initial survey of the region and preparation for GeMS/GSAOI observations we obtained Flamingos 2 (F2) \filter{JHKs} imaging of N79. 
The field of view is 6.1\arcmin\  in diameter with a 0.18\arcsec\ pixel scale. 
The observations were taken in a standard manner with dedicated sky frames due to the target region being crowded. Dithers were employed to account for bad pixels. 
For the \filter{J} band 25 frames of 60 seconds each,  for  the \filter{H} band 60 frames of 10 seconds each, and for the \filter{Ks} band 20 frames of 10 seconds each, respectively, were obtained. 

\begin{figure*} 
\begin{center}
\includegraphics[width=12cm]{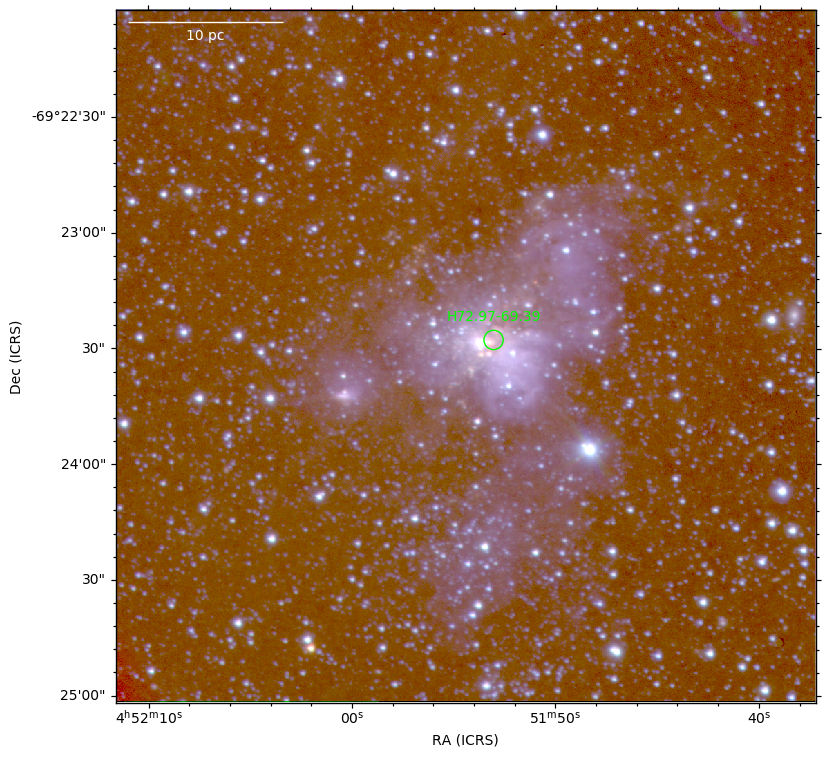} 
\caption{{  Flamingos 2} \filter{JHKs} color composite of N79 and its surroundings. Blue represents the \filter{J} band, green \filter{H}, and red \filter{$K_s$}. A logarithmic scale for all three channels has been used. The field of view shown is 3\arcmin$\times$3\arcmin , corresponding to 43pc$\times$43pc.  The whole nebulous region shown has previously been identified as IC2111. The location of \HH\ is marked with the green circle, the size (5\arcsec) of the Herschel beam.  } 
\label{col_F2} 
\end{center}
\end{figure*}

The data were reduced in a standard manner using the Gemini package within {\tt pyraf}. Dark frames were subtracted before flat fielding. 
For each science frame the adjacent sky frames were median combined and subtracted from the science frame. 
The sky subtracted science frames were then registered and combined rejecting outliers to remove cosmic rays and any remaining bad pixels. 
The circular field of view is vignetted by the peripheral wavefront sensor and for this paper we focus on the central 3\arcmin$\times$3\arcmin\  field of view. 
The trimmed color image representation of the F2 \filter{JHKs} data is shown in Fig.~\ref{col_F2}.

\subsection{GSAOI imaging} 
\filter{H} and \filter{Ks} band data of the central region of N79 were obtained on August 6 and October 23, 2018 with Gemini's GSAOI imager fed by the GeMS Multiconjugate Adaptive Optics system under program GS-2018B-Q-202, PI M. Andersen. 
The field of view of GSAOI  is 85\arcsec $\times$ 85\arcsec\ with a pixel scale of 20mas. 
Using five laser spots on the sky and three natural guide stars GeMS provide improved spatial resolution across the field of view. { The spatial resolution for the observations obtained here were 130mas for the \filter{Ks} band and 140mas in the \filter{H} band.  For comparison, the natural seeing in the \filter{H} and \filter{K} bands during the observations was 0.7\arcsec .
The seeing was rather uniform within a box 10\arcsec\ larger than the laser guide star configuration of 60\arcsec , but  deteriorates outside. }

For both filters individual exposures on source and on sky were obtained,  each with a length of 120 seconds. 
For the \filter{Ks} band a total of 10 frames were obtained for a total integration time of 20 minutes per pixel on-source. 
In the \filter{H} band 15 frames were obtained for a total exposure time of 30 minutes per pixel on-source. 

Only a fraction of the data requested were obtained due to variable weather and an earthquake. 
We thus did not obtain short exposures with GSAOI and the brightest few stars are moderately saturated in one of the filters. 
Consequently two stars are mildly saturated within the central cluster. For those two stars the F2 photometry will be used as discussed below. 

\begin{figure*}[ht]
\begin{center}
\includegraphics[width=15cm]{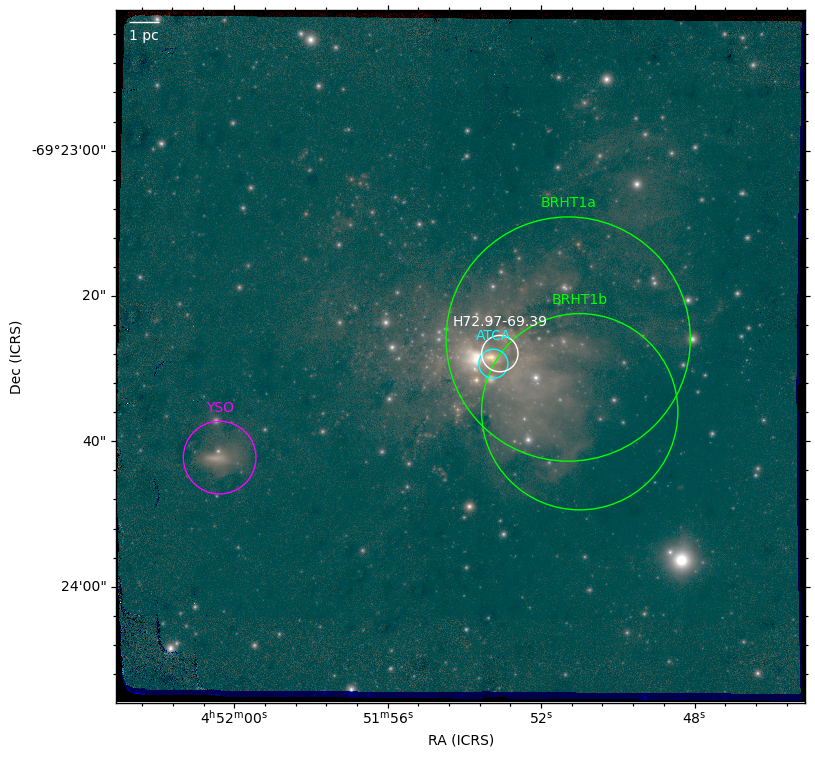} 
\caption{{ GeMS/GSAOI} \filter{HKs} color composite of N79 and its surroundings. The field of view is 90\arcsec$\times$90\arcsec { (22.5pc$\times22.5pc$}. The green channel is created as an average of the \filter{H} and \filter{Ks} images. The stretch is a square root stretch. Shown are the locations of several known regions, including the \citet{bhatia} double clusters (shown as the two large { green} circles), the ATCA source (small cyan circle), and \HH\  in green. To the east, encircled in magenta is located a YSO \citep{gruendl} likely unrelated to \HH , as discussed in the text. } 
\label{col_GSAOI} 
\end{center}
\end{figure*}

The images were reduced in a standard manner using the Gemini data reduction package within the \texttt{pyraf} environment. 
Dark current was subtracted before the individual frames were flat fielded. 
The sky frames were median combined and the resulting sky frame was subtracted from the science frames. 
The frames were distortion corrected to a single reference frame using stars in common and using the dedicated software package {\tt disco-stu}. 
All distortion corrected frames were combined into the final \filter{H} and \filter{Ks} frames. 
After averaging the distortion corrected frames for each filter we block averaged the frames by a factor of two resulting in a final pixel scale in the data of 40mas. 
Fig.~\ref{col_GSAOI} shows the region observed with GSAOI where { several} previously identified regions are { marked}. 

\subsection{Source detection}  
The focus of this study is on the high spatial resolution GSAOI data. However, we will further use the Flamingos 2 data to calibrate the GSAOI data and to provide photometry for a few sources that are saturated in the GSAOI images. 

Sources in the GSAOI images were detected in the \filter{Ks} band image using {\tt daofind} in the {\tt pyraf} environment. 
Sources were identified in the \filter{Ks} band and that list was then used as the input list for the \filter{H} band image. 
Sources were identified down to a threshold of three times the standard deviation of the background. 
A source was considered detected in both bands if the position after photometry was the same in the two bands to within 1 block average pixel of 40 mas.

\subsection{Photometry and photometric calibration}
\begin{figure}[ht]
    \centering
    \includegraphics[width=8.5cm]{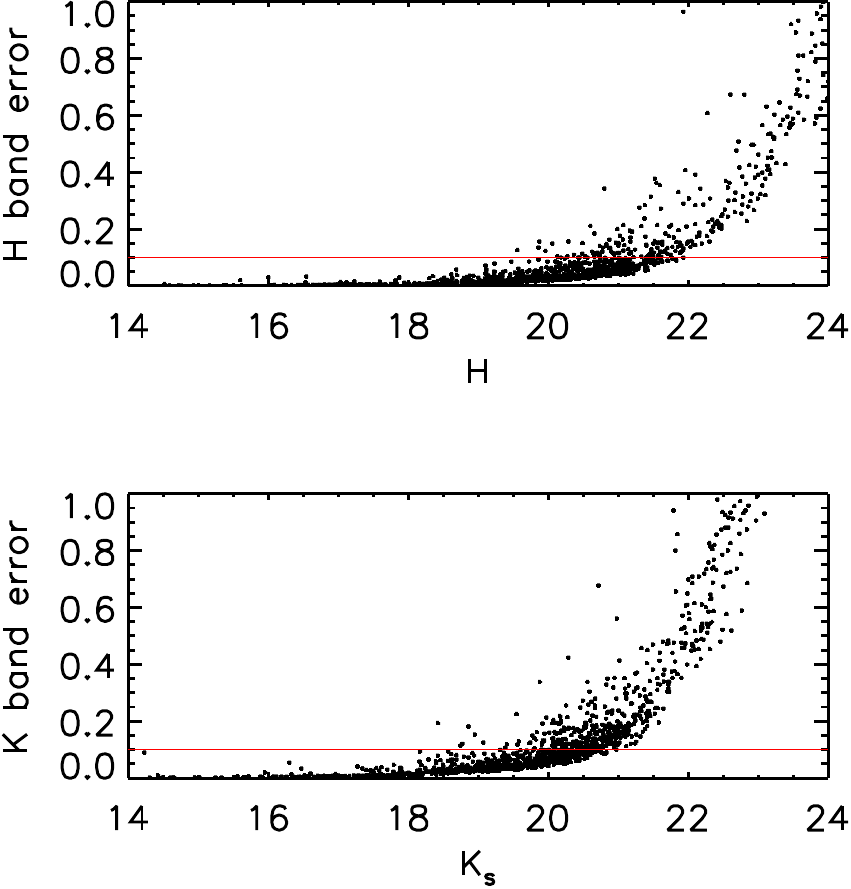}
    
    \caption{The derived magnitudes and photometric errors for the stars in the  GSAOI \filter{H} band (top) and \filter{Ks} band (bottom). 
    A magnitude error of 0.1 mag is highlighted by the horizontal line in both diagrams. }
    \label{mag_merr}
\end{figure}

Photometry was performed in a standard manner. 
An aperture with a radius of 3.5 pixels (140mas),  equal to the full width at half maximum was used.  The sky was estimated independently for each source in an annulus covering 0\farcs8 to 1\farcs2 from the source center. 
In both the \filter{H} and \filter{Ks} band images we used the \filter{Ks} band source list and allowed for a recentering of each source.  
The photometric error as a function of magnitude is shown in Fig.~\ref{mag_merr} for both filters. 

For the results and analysis we limit the sample to objects with a magnitude error less than 0.1 mag in both filters. A total of 965 sources were detected, given these criteria. 

The GSAOI field of view is relatively small and there are few sources in common with 2MASS to calibrate with. 
Instead we use the F2 data which covers a larger field and due to shallower observations have a better overlap in magnitudes with 2MASS. 
The transformation was based on 26 stars in common between 2MASS and the short exposure F2 data. 
All stars used had a photometric error in each of the 2MASS bands of less than 0.1 mag and were visually inspected to ensure they were not unresolved binaries in the 2MASS data or affected by nebulosity. 
Subsequently a list of well exposed stars in common between the F2 and GSAOI datasets were used to calibrate the GSAOI data. 


Several stars are saturated in one of the GSAOI bands. For those stars we use the Flamingos 2 photometry. 
To limit the flux contamination from other nearby sources and nebulosity we have derived the photometry for the stars using an aperture of 1.5 pixel radius (corresponding to 0\farcs 27 for F2). 

{  The source list is presented in Table~\ref{phot_table}, where the photometry in the 2MASS system is provided. 
 \begin{table*}
\caption{Sources identified within the GSAOI field with photometric errors below 0.1mag. Five sources are saturated in one or both GSAOI bands and they have been replaced with Flamingos 2 photometry. The full list is available in the online version.}
\begin{tabular}{cccccccc}
Source ID & RA & DEC & \filter{H}& error \filter{H} & \filter{Ks} & error \filter{Ks} & Instrument \\
\hline
       0   &    72.983836&      -69.389467&20.2& 0.04 &20.0 &0.09 &GSAOI\\
       1    &   72.963730&      -69.389388&20.1& 0.02 &19.9 &0.06 &GSAOI\\
       2    &   72.969217&      -69.389391&20.7& 0.06 &19.9 &0.08 &GSAOI\\
       3    &   72.944780&      -69.389353&20.4& 0.03 &20.3 &0.09 &GSAOI\\
       4    &   72.986876&      -69.389369&18.2& 0.00 &18.0 &0.01 &GSAOI\\
       5    &   72.981919&      -69.379712&20.3& 0.03 &20.1 &0.08 &GSAOI\\
       6    &   72.943188&      -69.403334&16.0& 0.00 &15.9 &0.00 &GSAOI\\
       7    &   72.987428&      -69.389264&19.8& 0.02 &19.6 &0.05 &GSAOI\\
       8    &   72.964797&      -69.389233&19.2& 0.01 &18.3 &0.01 &GSAOI\\
       9    &   72.964706&      -69.389204&19.3& 0.01 &18.4 &0.01 &GSAOI\\
\hline
\label{phot_table}
\end{tabular}
\end{table*}
}

\subsection{Archival Herschel and ALMA data} 
{ Previous work has examined  the dust and gas content of \HH\ using {\it Herschel} and ALMA data \citep{ochsendorf,nayak19}. 
To place the GeMS/GSAOI data in context we utilize some of these data as they were presented in the literature but here we briefly describe the used data. 

\HH\ was identified as a far-infrared point source in \citet{meixner} using PACS and SPIRE imaging from {\it Herschel}. These data were combined with {\it Spitzer} and {\it WISE} data in \citet{ochsendorf} to estimate the total luminosity of \HH . 
Note that the spatial resolution for the far-infrared observations is 5.5\arcsec at 60 $\mu$m and worse at longer wavelengths. 

\citet{nayak19} presented ALMA observations of multiple molecular species and the H30$\alpha$ recombination line. Here we use their $^{12}CO(2-1)$, $^{13}CO(2-1)$, $SO(6-5)$, and H30$\alpha$ data. 
Observations were obtained in both compact and extended mode resulting in a spatial resolution ranging from 0.239\arcsec$\times$0.160\arcsec\ for $SO(6-5)$ to 0.331\arcsec$\times$0.224\arcsec\ for $^{13}CO(2-1)$. 
Due to the compact configuration the maximum recoverable scale is 17.8\arcsec . 

}
\section{Results}

\subsection{The environment of \HH}
{ Fig.~\ref{col_F2} shows the Flamingos 2 color image where the IC2111 HII region is evident.} In Fig.~\ref{col_GSAOI} some of the previously identified objects and regions near \HH\  are marked. These include the  clusters BRHT1a and BRHT1b suggested to be double clusters \citep{bhatia},  and \HH\ itself which coincides with the ATCA detected source { per the coordinates provided in \citet{indebetouw}}.  
The ATCA source spatially aligns with the {\it Herschel} source \citep{meixner} as well as the $\mathrm{H30\alpha}$ source identified in \citet{nayak19} within a few arcseconds, i.e. the expected accuracy of the lower--resolution {\it Herschel} studies. These sources coincide with a bright nebulous region in the near-infrared with a bright point source associated with it.

Some 25\arcsec\  to the north--east of \HH\ there is a cluster of reddened objects. The lack of strong nebulosity could suggest it is a slightly older population than the one associated with the \HH\ region. 

Fig.~\ref{col_F2} shows that the diffuse nebulosity extends further to the south of the two clusters and based on the color--magnitude diagrams presented below the whole region is extincted by the surrounding dust. 
This is not surprising since  \HH\ is part of the N79 star forming complex and as such we do expect several star formation sites within the region. 
Here we focus on the stellar content in the vicinity of \HH , and based on the radial surface mass density profile in Fig.~\ref{radial} we focus on central 8\arcsec\ radius region.

Further to the east of the \HH\  is located one extended object marked in Fig.~\ref{col_GSAOI}. It was identified in \citet{gruendl} as a YSO based on the near-infrared and mid-infrared colors from 2MASS and Spitzer imaging. We note that there is both a point--like object and an extended and elongated nebulous region below it, with a length of roughly 2\arcsec , similar to the spatial resolution of 2MASS so the source has appeared as a point source in the study of \citet{gruendl}.  Due to the large  projected distance of the YSO from \HH\ (10 pc)  this source is  not  likely to be directly related to \HH .

\begin{figure*}[ht]
    \centering
    \includegraphics[width=14cm]{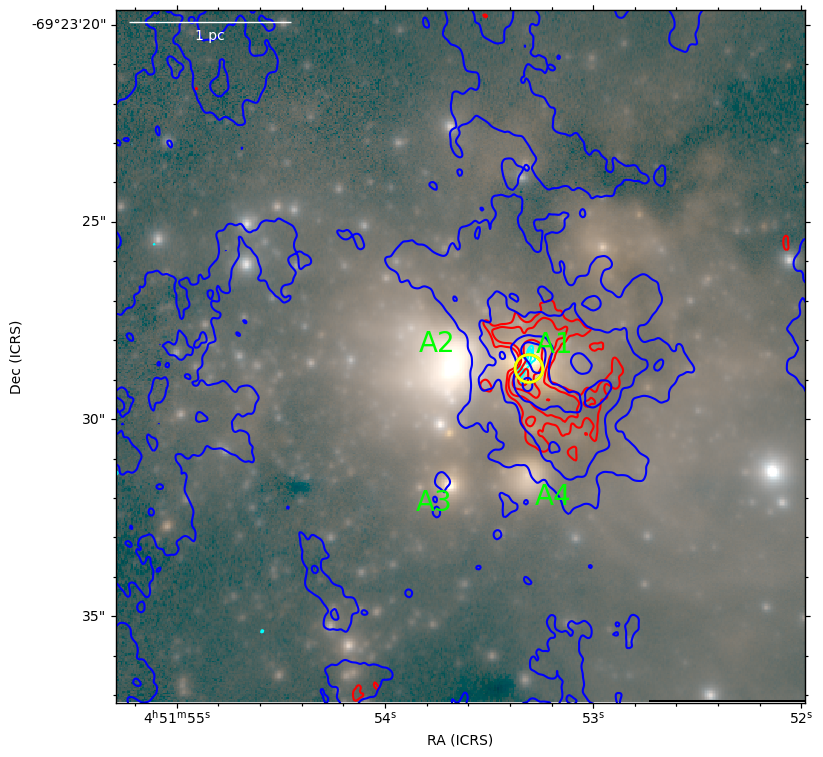}
    
    \caption{The central region around \HH\ . Several filamentary dark lanes are present obscuring parts of the stellar content even at near-infrared wavelengths. 
    Overplotted as contours are the SO integrated emission (cyan), the $^{13}CO$ integrated emission in red, and the $^{12}CO$ integrated emission in blue \citep{nayak19}. { The location of the H30$\alpha$ source is further marked as a yellow circle. }
    The spatial resolution of the ALMA 1.3mm observations is 0.3\arcsec$\times$ 0.2\arcsec . 
    { Several bright sources associated with nebulosity have been marked as A1 (the ATCA source), A2 (the approximate cluster center, A3, and A4 )}.}
    \label{GSAOI_zoom}
\end{figure*}

Figure~\ref{GSAOI_zoom} shows in detail the GSAOI image of  the region around \HH\ with contours of ALMA detected emission in several molecular tracers shown.  
The higher spatial resolution of the GSAOI images compared to the previous near-infrared imaging shows that although there is a bright star associated with extended nebulosity next to the H30$\alpha$ source, they are offset by $\sim$ 0.2\arcsec, which, however, is  comparable to the ALMA beam size and pointing accuracy of ALMA. 
The integrated SO emission is extending almost perpendicular to the nebulosity. 
Whereas the $SO$ emission is compact, the $^{13}CO$ emission is more extended with a peak of the integrated emission near source A1. 
The $^{13}CO$ emission traces more quiescent gas and is thus showing the extent of molecular material around \HH\ whereas $SO$ is tracing either high density or shocked material. 


In the vicinity of \HH\ there are several dark filamentary lanes seen extending from the south and west of \HH . 
The dark lanes are loosely associated with the emission in $^{13}CO$ and in particular  $^{12}CO$.   Since they appear dark relative to the extended nebulosity they are in the foreground but based on their spatial alignment appears to be associated with \HH .

Several other bright infrared sources are evident in the vicinity of \HH . These include one to the East (2\arcsec , or 0.5pc, source A2) associated with strong nebulosity and another nebulous source 2\farcs 9 (0.7pc) to the south (source A4). 
Some 3\farcs 8 (0.9pc) to the South-East of \HH\  is located another bright object associated with nebulosity (source A3). 
The fact that these bright objects are associated with nebulosity and their proximity to \HH\ suggests that they are young and most likely connected to the recent star formation in the region.

\subsection{The center of \HH\  in the near-infrared}
{ We have determined the near--infrared center of \HH\ by measuring the stellar number density profile in both RA and DEC and defined the center as the  peak in each direction}. 
The sample has further been limited to objects included in the mass estimate described in Section 4, with stellar masses above 10 M$_\odot$. We have examined if there is any dependence on the center if potential foreground objects are excluded using an extinction cut (using objects with A$_{Ks}$ larger than 0.75 only) but the center does not change. 
The derived peak is located to the east of \HH\ by 0.5pc, close to the bright source marked as A2 in Fig.~\ref{GSAOI_zoom}

\subsection{Luminosity functions and color-magnitude diagrams}
\begin{figure}[ht]
    \centering
    \includegraphics[width=7cm]{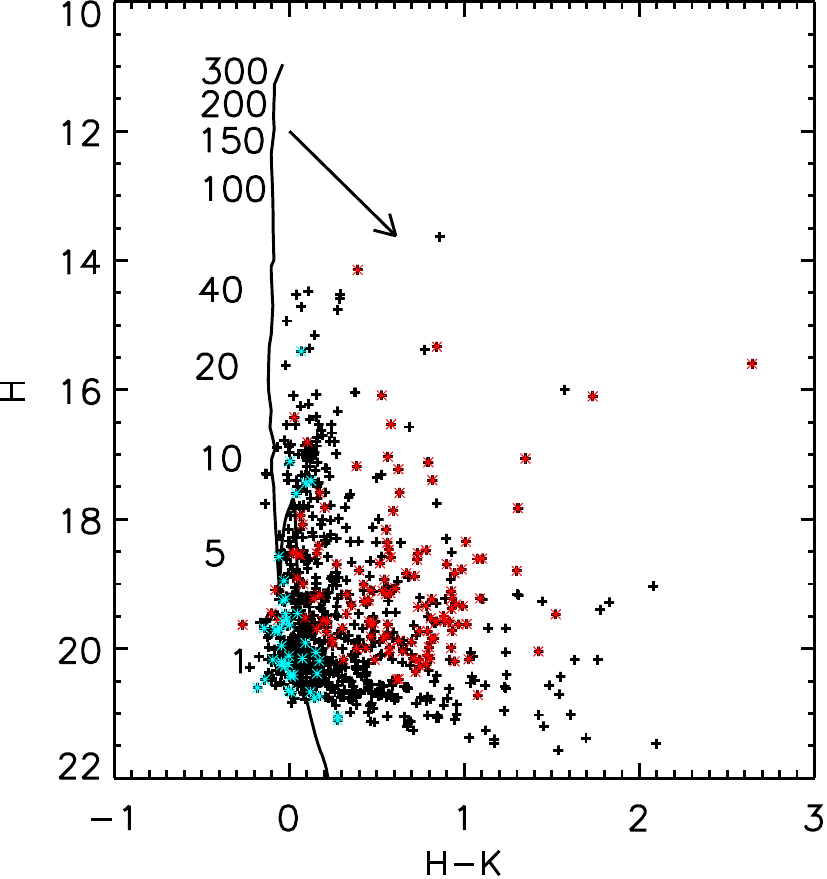}
    \includegraphics[width=7cm]{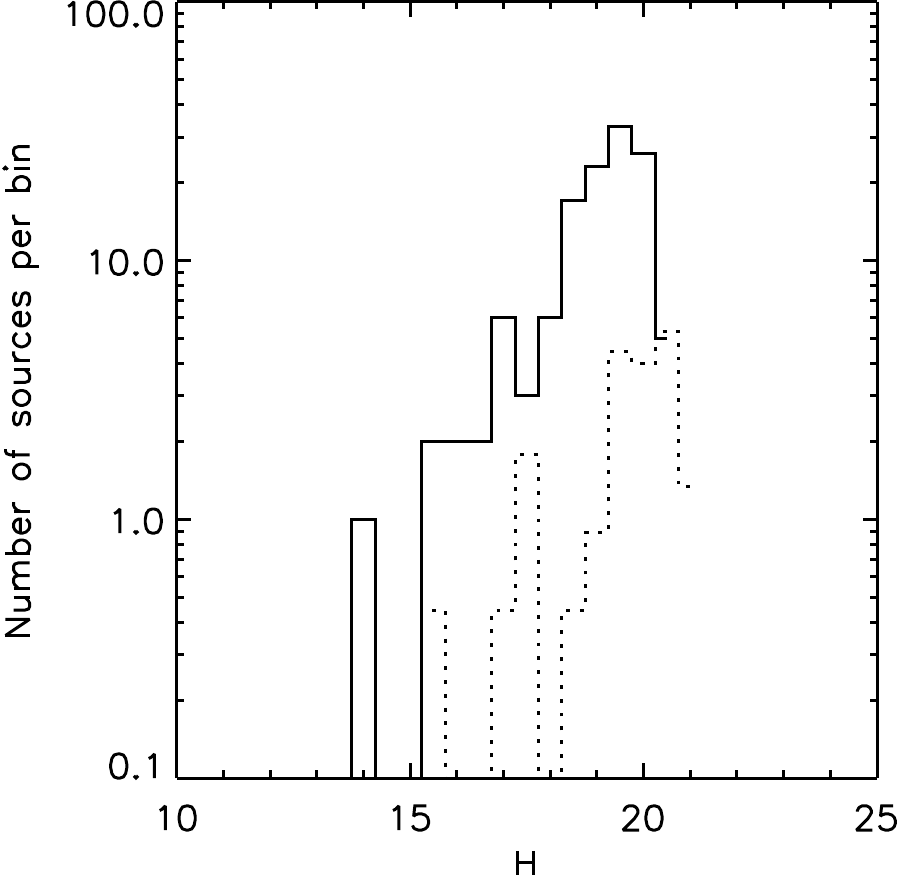}
    
    \caption{Top: \filter{H}-\filter{K} versus \filter{H} band color magnitude diagram for the point sources in the  GSAOI field of view. Red points are those within 200 pixel radius (2 pc) of the central parts of \HH , cyan in the control field to the South. Overplotted is a 1 Myr isochrone from the PARSEC calculations \citep{tang}. 
    { The numbers indicate the initial mass for a star with that \filter{H-K} color and \filter{H} band magnitude.} Note that the main-sequence to pre-main sequence transition is not monotonic as discussed in the text. 
    Bottom: The luminosity function for stars in the cluster region (solid histogram) and the control region (dashed histogram). }
    \label{CMD}
\end{figure}

\begin{figure}[ht]
    \centering
    \includegraphics[width=7cm]{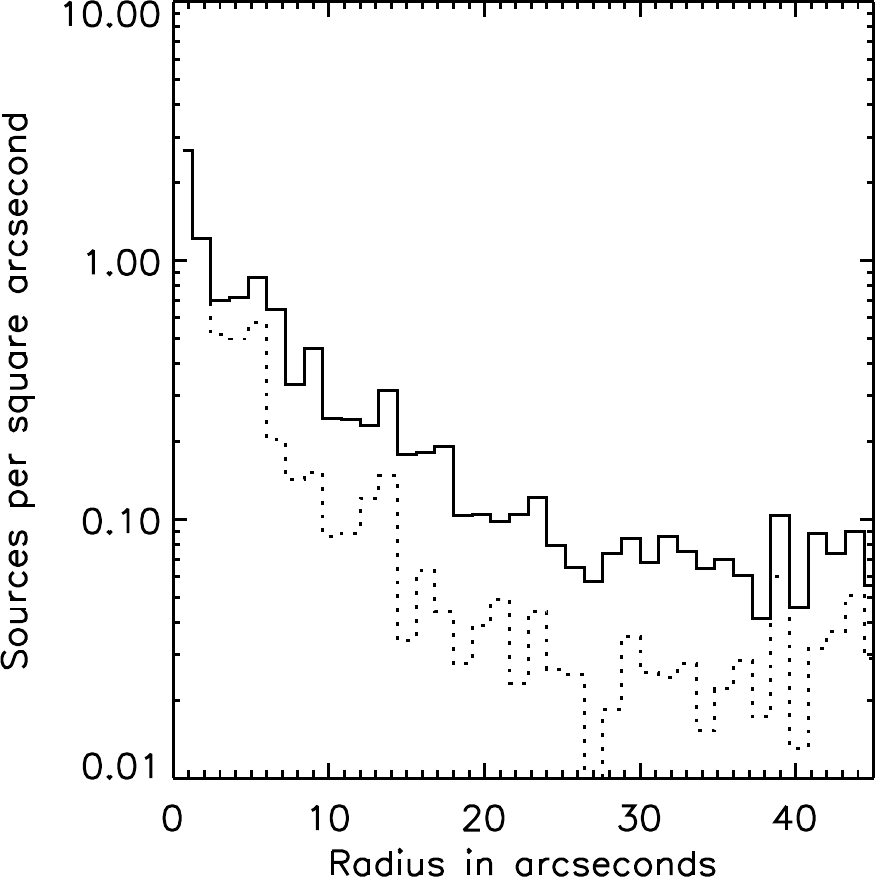}

    \caption{The radial surface density profile for sources in the region around \HH\ . The solid-lined histogram is for all detected sources with a photometric error less than 0.1 mag in each band and the dashed-lined histogram is for sources brighter than an observed magnitude of  \filter{H}$=19.5$, where the completeness is less affected by differential extinction.}
    \label{radial}
\end{figure}

Fig.~\ref{CMD} shows the color-magnitude diagram across the GSAOI field of view for the sample of objects with photometric errors less than 0.1 mag in each band. 
Objects within 200 pixels radius (2 pc) are marked in red. 
The radius of 2 pc is based on Fig.~\ref{radial} where the stellar surface density reaches 10\%\ of the peak value.
{ There is a large range of reddening values for the region, from objects close to the isochrone and then extending to \filter{H-K} of 2 mag. 
In the  near-infrared color-magnitude diagrams the isochrone for higher mass stars is almost vertical and there is little dependence as a function of mass. However, the extinction is derived by de-reddening the object to the isochrone as described below.} 
Not only the sources in the vicinity of \HH\  are affected by reddening, also across the region is this true as can be seen by some of the red sources in the control field. 

The large range of reddening means there is no well-defined locus of the cluster members but there is a large over-density over the control field, 132 sources in the cluster region but only 59 in the control field (which covers four times the area). 
Further, the field star counts tend to be at fainter magnitudes where the nebulosity associated with \HH\  limits the detection of faint stars. 
This is seen in the luminosity functions for the two regions in Fig.~\ref{CMD}. 
The cluster field has a number of bright sources not present  in the control field but with a relatively larger number of sources at faint magnitudes. 
This is consistent with the control field being populated predominantly with an older foreground  population. 

Fig.~\ref{radial} shows the radial profile for the region around \HH , centered as noted in Fig.~\ref{GSAOI_zoom}. 
The profile peaks strongly at the location of \HH\ and declines gradually towards a background level outside 20 \arcsec . 
Limiting the sample to sources brighter than \filter{H}$=19.5$ shows the same number of detected sources in the central region as the full sample but then a steeper decline with radius suggesting extinction has a stronger effect in the center. 

\section{Analysis} 

\subsection{De-reddening and stellar mass estimates}
In order to derive the stellar masses an extinction law and a stellar isochrone are needed. 
For the infrared it has been found that the extinction curve in the LMC does not differ substantially from the Milky Way \citep[e.g.][]{koornneef82} and thus we adopt the parameterisation of \citet{cardelli}. We used an $R_V$ of 3.4, the average $R_V$ reported by \citet{gordon} for their LMC stellar sample. 
For the stellar isochrone we adopt the PARSEC models which covers young ages as well as low metallicity, appropriate for the LMC. 
We use a distance modulus of 18.48, corresponding to a distance of 49.6 kpc \citep{dist_LMC}. 
{ For each mass in the model a bolometric magnitude and effective temperature are provided which in turn provide the predicted \filter{H} and \filter{Ks} band magnitudes. From these, together with the reddening law, each object can be slided along the reddening vector in the color-magnitude diagram to the isochrone and the mass for the object can be determined together with a measure of the extinction. 
The isochrone has a finite resolution and linear interpolation is performed from the two closest points on the isochrone. 
}

The stellar isochrones vary with age, especially for the lower-mass content. 
The mass where the pre-main sequence to main sequence transition happens is age dependent and is lower for an older population. 
The transition results in a brightening of the star which translates into the non-monotonicity seen in the isochrone in Fig.~\ref{CMD} for the mass range 5-10 M$_\odot$. 
For a 1 Myr old population the transition happens at a mass of $\sim$ 3 M$_\odot$ where the stars can be as bright as a 9 M$_\odot$ main sequence star in the \filter{H} band. 
For a 2 Myr old population the transition is at 2.5 M$_\odot$. 
This does introduce a degeneracy for mass determination over the range where the relationship between stellar mass and its luminosity is not monotonic and we will explore the effects of that below when estimating the cluster mass.

With only the  \filter{H} and \filter{Ks} filters, identifying objects with a reddening and a  near-infrared excess due to a circumstellar disk is difficult since the reddening and presence of a disk would be degenerate. 
The disk fraction in massive clusters is still poorly known, especially for high mass stars. \citet{stolte} found a modest disk fraction of less than 10\% in the Arches and Quintuplet clusters using \filter{L} band observations. 
Since \filter{L} band observations are more sensitive to  circumstellar disks than \filter{Ks} band  observations \citep{haisch} we would expect to identify even fewer disks in a near-infrared survey than was the case in \citet{stolte}. 
In addition, there is even fewer indications of disks in the near-infrared for massive stars, as probed here, making the possible contamination even smaller.
Thus, even though \HH\  is likely younger than the Quintuplet and Arches we would still expect the disk fraction for the intermediate-- and high-- mass stars to be small and not affect the derived masses.

\begin{figure}[ht]
    \centering
    \includegraphics[width=7cm]{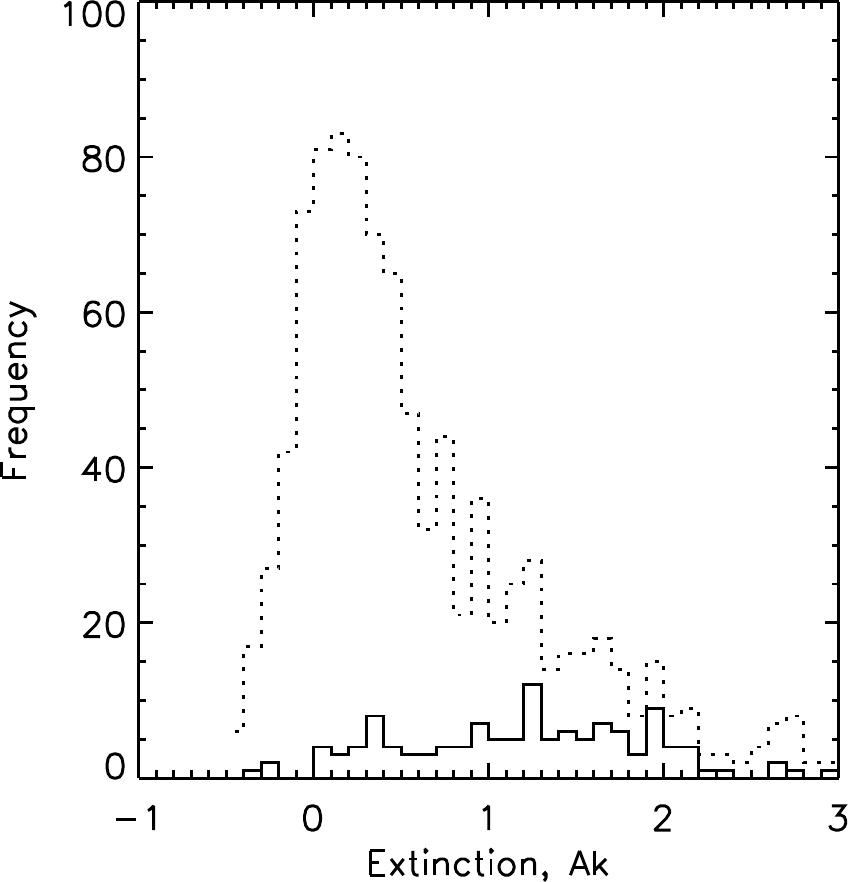}

    \caption{The extinction derived using the 1 Myr isochrone for all stars within the GSAOI field of view (short-dashed histogram) and for the region around \HH\  (solid-lined histogram).}
    \label{ext}
\end{figure}

Fig.~\ref{ext} shows the distribution of derived extinction values for the GSAOI field as a whole and the \HH\ region. 
The distribution of extinction is  different for the field as a whole and for \HH . Whereas the field in general has a distribution that peaks at low extinction, consistent with the foreground extinction towards the LMC, the distribution for \HH\  has a much broader peak at $\mathrm{A_K}=1.2$ and a median value of 1.25. 
However, the extinction distribution is likely skewed towards less extincted objects due to the limited depth of the observations. 
If instead the sample is limited to the brighter stars with a mass above 10 M$_\odot$ (64 objects) the median extinction is $A_{Ks}=1.7$.  
Assuming there is no preference for higher mass stars to be more deeply embedded than lower mass stars this would suggest a relatively fewer low-mass stars to be detected. 
In addition, if an unbiased measure of an Initial Mass Function is needed, an extinction limited sample needs to be constructed \citep[e.g.][]{andersen09}. 

\subsection{The total stellar mass of \HH\  }
{ The total stellar mass of \HH\  is a key parameter to determine if the region will evolve into a R136 type region in the future. 
Based on the de--reddened \filter{H} band magnitudes derived in the previous section we can obtain several estimates of the total stellar mass. 
The most straight forward is to calculate the mass of the detected stars. 
Using the masses determined above, a total mass of 2466 M$_\odot$ is found for a 1 Myr isochrone assuming a metallicity of $-0.5$ dex, based on the metallicity maps for the LMC in \citet{choudhury18}. 
However, this does not take field star contamination into account. 
We use the control field in the North-East corner to estimate the field star contamination to the mass. 
{ The radius is 300 binned pixels compared to the 200 binned pixel radius for the science region for better statistics. }
Since field stars within the cluster would have been assigned a mass assuming the 1 Myr isochrone we use the same isochrone to estimate masses for the control field stars. 
The total mass of stars in the control field is 141 M$_\odot$ but this is for a larger area. Taking this into account the field star corrected directly detected mass of \HH\  is 2403 M$_\odot$. 

However, the finite depth of the imaging means only a fraction of the stellar content is determined. 
Extrapolating to low masses can provide an estimate of the full stellar content. 
Assuming a log-normal IMF below 1 M$_\odot$ and a power-law above using the prescription in \citet{chabrier} the full stellar mass is extrapolated. 
Although there is a large ongoing debate if the IMF is variable as a function of environment, there is little to indicate that this is the case for present day clusters in the Milky Way or Large Magellanic Clouds \citep{andersen09,bastian,dario09,andersen17a} with the possible exception of the clusters near the Galactic center \citep{hosek}, which is a very different environment than N79.

For the detected stellar mass we restrict the sample to 10 M$_\odot$ and above to avoid issues with any inflections in the mass--luminosity relation and incompleteness. 
{ We detect a total mass of 2186 M$_\odot$ in stars more massive than 10 M$_\odot$. 
However, this includes potential contamination from field objects. Indeed, correcting for the field star contamination this is reduced moderately to 2144 M$_\odot$. 
Using the field star corrected mass estimate to extrapolate the IMF to low masses the total mass is found to be 10951 M$_\odot$. }

We have performed the same calculation with  0.5 and 1.6 Myr isochrones. The total extrapolated masses are 12602 M$_\odot$, and 9716 M$_\odot$, respectively.  
Due to cluster being deeply embedded, the presence of molecular filaments and shocks indicative of young stars we prefer a younger age and thus the mass of \HH\ is expected to be above $10^4$ M$_\odot$ within 2 pc radius. 
}
\subsection{The immediate environment of the ALMA peak} 
As shown in Section 3, the peak of the stellar content of the \HH\ region is centered next to A2 in Fig.~\ref{GSAOI_zoom}, to the west of the ALMA peak location. 
The peak of the stellar distribution shows less emission from $^{12}CO$ and $^{13}CO$  as well as lower extinction than in the vicinity of the ALMA peak and is thus likely an older star formation event than the ALMA region of \HH .

We have investigated the immediate stellar content around the \HH\ ALMA source, within a radius of 1.6\arcsec .  
A total of six sources are found within this region, the least massive of which is 17 M$_\odot$ (independent of the age assumed to be 0.5, 1, or 1.6 Myr). 
The total stellar mass is found to be 428 M$_\odot$. 
The average surface mass density in the region is 53 M$_\odot$ arcsec$^{-2}$. Since the lowest mass object is 17 M$_\odot$ one can directly compare with the surface mass density in the center of R136 \citep{brandl96}. Their single conjugate adaptive optics study had a spatial resolution similar to this study. 
They determined a surface mass density within the central 0\farcs 4 to be $\sim$ 1200 M$_\odot$ arcsec$^{-2}$ for stars more massive than 15 M$_\odot$.

\subsection{Is \HH\  a forming SSC?}
{ We find that \HH\ contains a population of massive stars, but, compared to a region like R136 the total mass of the region is relatively modest.} 
The 2 pc radius adopted is similar to the half-mass radius of R136 within the 30 Dor complex \citep[1.7 pc][]{hunter95,andersen09}. 
The extrapolated total mass of R136 was estimated in \citet{andersen09} to be $10^5$ M$_\odot$, and thus the mass within the half-mass radius $50,000$ M$_\odot$, more than a factor of five larger than that of \HH\ , where the mass within 1.7 pc is found to be 7200 M$_\odot$. 

{ However, whereas R136 is largely free of gas, there is still material present for potential further star formation in \HH . 
To gauge the amount available we use the extinction determined for the individual stars to estimate the total amount of dust and hence gas within \HH . 
We use the average extinction derived for the stars above 10 M$_\odot$, similar as above to estimate the total stellar mass. 
The extinction is converted into a gas mass estimate based on the relations between $A_V$ and dust surface mass density and a gas to dust ratio of 380  determined by \cite{duval} for the LMC. 
{ The conversion of measured extinction to dust mass relies on the extinction law being standard in the near-infrared for \HH . 
If the dust size distribution is different due to e.g. grain growth this will affect the derived extinction values and hence the dust mass.  The dust to gas ratio is different for the LMC than the Milky Way and depends on metallicity. 
Any change in the gas to dust ratio locally will linearly affect the derived dust mass.}
There has recently been claims of large reservoir of CO dark gas, most notably in the 30 Dor region which could suggest we will underestimate the total gas mass \citep{chevance}. 
However, as discussed in \citet{chevance} this is mainly observed at low extinction values, typically $A_V$ of less than three. In the case of \HH , the average extinction is $A_V > 10$. 

We have adopted the average extinction to hydrogen column density conversion from \citet{duval}. 
Converting the derived near-infrared extinction to the extinction in the \filter{V} band using the \citet{cardelli} extinction law we obtained a total gas mass 6580 M$_\odot$ within a radius of 2 pc. 
This is based on the average  extinction of the sight lines probed by the embedded stars (126 stars). 
The SO traced gas will likely add to the total molecular gas estimate. 
In addition, the $^{13}CO$  emission was highly filamentary \citep{nayak19} and will not be captured well by the extinction mapping. 
Relying only on the extinction mapping as a lower limit to the gas mass. 
Adding the gas mass determined above to the stellar mass gives a total mass of 17318 M$_\odot$ assuming a 1 Myr isochrone for the stellar content and extrapolating the IMF to the full stellar range. }
Given the stellar content within 2 pc of the central region of \HH , we find that the cluster is a Young Massive Cluster, based on the criteria in \citet{portegieszwart} of being more massive than $10^4$ M$_\odot$ and younger than 100 Myr.   
From the stellar density profile it is clear that the majority of mass is within the 2 pc radius, with an increase of the total mass of 30\%\ for a radius of 3 pc. 
Even with the gas mass added to the  stellar mass, \HH\ is still lower mass than R 136 for the same spatial coverage. 
Further, most likely not all the gas would be turned into stars and thus only a fraction would be added to the total final stellar mass. 
{ However, as the extinction distribution shows, a substantial fraction of the sightlines probe gas with an extinction higher than $A_{Ks}=0.8$ which is suggested the threshold for star formation in the Galaxy \citep{lada2010}}. 
All of this places a likely total final mass for \HH\  closer to 15000 M$_\odot$ which would make it a higher mass YMC. 
Based on the large amount of gas still in the region, and that the cluster is deeply embedded, \HH\ is expected to be a much younger YMC than  R136 and it thus provides a unique opportunity to study massive star cluster formation in a metal-poor environment.

\section{Summary} 
We have performed Flamingos 2 and GeMS/GSAOI imaging of the forming potential Super Star Cluster associated with \HH\  in the Large Magellanic Cloud. 
We find a near-infrared embedded cluster with a characteristic radius of $\sim$2 pc,
with a large range of reddening compared to the adjacent region. 
The cluster includes several bright, highly reddened sources associated 
with nebulosity. The total stellar mass of the 2 pc radius region around 
\HH\  is at least  $10^4$ M$_\odot$ based on the intermediate-- and high-- mass content, extrapolated for the whole stellar mass range, adopting a 1 Myr isochrone. 
The central mass surface density of the cluster is found to be smaller than that of R136, although it appears that \HH\ is still sub-clustered based on the existence of different stellar and molecular peaks. 
Thus, \HH\ coincides with a forming Young Massive Cluster,  offering an opportunity to study the formation of a sub-solar metallicity Young Massive Cluster.

\acknowledgements{{ We thank the referee and editor for suggestions and comments that improved the manuscript.} Based on observations through programs 2019A-DD-103, and  2018B-Q-202, obtained at the international Gemini Observatory, a program of NOIRLab, which is managed by the Association of Universities for Research in Astronomy (AURA) under a cooperative agreement with the National Science Foundation. on behalf of the Gemini Observatory partnership: the National Science Foundation (United States), National Research Council (Canada), Agencia Nacional de Investigaci\'{o}n y Desarrollo (Chile), Ministerio de Ciencia, Tecnolog\'{i}a e Innovaci\'{o}n (Argentina), Minist\'{e}rio da Ci\^{e}ncia, Tecnologia, Inova\c{c}\~{o}es e Comunica\c{c}\~{o}es (Brazil), and Korea Astronomy and Space Science Institute (Republic of Korea). { Nayak and Meixner were supported by NSF grant AST-1312902, NNX15AF17G, HST-GO-14689.005-A, and SOFIA: SOF-04-0101. Nayak was additionally supported by STScI Director’s Discretionary Fund. Hirschauer  and Meixner acknowledge support from NASA grant NNX14AN06G.}}

\end{document}